\documentclass{emulateapj}

\usepackage{psfig}
\usepackage{amssymb}
\usepackage{graphicx}

\newcommand{\lSect}[1]{{\label{sec:#1}}}
\newcommand{\lFig}[1]{{\label{fig:#1}}}
\newcommand{\lEq}[1]{{\label{eq:#1}}}

\def\gtaprx {\lower .1ex\hbox{\rlap{\raise .6ex\hbox{\hskip .3ex
	{\ifmmode{\scriptscriptstyle >}\else
		{$\scriptscriptstyle >$}\fi}}}
	\kern -.4ex{\ifmmode{\scriptscriptstyle \sim}\else
		{$\scriptscriptstyle\sim$}\fi}}}
\def\ltaprx {\lower .1ex\hbox{\rlap{\raise .6ex\hbox{\hskip .3ex
	{\ifmmode{\scriptscriptstyle <}\else
		{$\scriptscriptstyle <$}\fi}}}
	\kern -.4ex{\ifmmode{\scriptscriptstyle \sim}\else
		{$\scriptscriptstyle\sim$}\fi}}}
\newcommand{\FIGFF}[2]{{\ref{fig:#2}{#1}}}

\newcommand{\FIG}[2]{{Fig.~\FIGFF{#1}{#2}}}
\newcommand{\Fig}[1]{{\FIG{}{#1}}}

\newcommand{\Sectff}[1]{{\ref{sec:#1}}}
\newcommand{\Sect}[1]{{\S~\Sectff{#1}}}

\newcommand{\Eqref}[1]{{\ref{eq:#1}}}
\newcommand{\Eqff}[1]{{(\Eqref{#1})}}
\newcommand{\EQ}[1]{{Equation~\Eqff{#1}}}

\newcommand{\Eq}[1]{{eq.~\Eqff{#1}}}

\begin{document}


\title{Type~Ia Supernova: Burning and Detonation in the Distributed Regime}

\author{S. E. Woosley\altaffilmark{1}}

\altaffiltext{1}{Department of Astronomy and Astrophysics,
University of California, Santa Cruz, CA 95064; woosley@ucolick.org}

\begin{abstract} 
A simple, semi-analytic representation is developed for nuclear
burning in Type Ia supernovae in the special case where turbulent
eddies completely disrupt the flame. The speed and width of the
``distributed'' flame front are derived.  For the conditions
considered, the burning front can be considered as a turbulent flame
brush composed of corrugated sheets of well-mixed flames. These flames
are assumed to have a quasi-steady-state structure similar to the
laminar flame structure, but controlled by turbulent
diffusion. Detonations cannot appear in the system as long as
distributed flames are still quasi-steady-state, but this condition is
violated when the distributed flame width becomes comparable to the
size of largest turbulent eddies. When this happens, a transition to
detonation may occur. For current best estimates of the 
turbulent energy, the most likely density for the transition to
detonation is in the range $0.5 - 1.5 \times 10^7$ g cm$^{-3}$.
\end{abstract}


\section{INTRODUCTION}
\lSect{intro}

One of the greatest uncertainties in how a Chandrasekhar mass white
dwarf explodes as a Type Ia supernova is whether and how an initially
subsonic burning front, a deflagration, makes a transition to a
supersonic detonation.  A related question is the characteristics of
nuclear burning in a medium where turbulence has become so strong that
hot ash and cold fuel can be co-mingled before burning. The conditions
for the latter, known as ``burning in the distributed regime'',
have long been known to both the combustion and astronomical
communities \citep{Pop87,Nie97a,Kho97,Nie97,Pet00}. When the laminar
flame grows thick enough and the turbulent intensity great enough,
the ``Gibson'' length, that length scale which turns over due to
turbulent eddies as fast as a laminar flame crosses it, becomes
smaller than the flame thickness itself.  That is,
\begin{equation}
\frac{L_{\rm Gib}}{v_{\rm Lam}} \gtaprx \tau_{\rm turb}(L_{\rm Gib}).
\end{equation}

Approximate conditions for entering the distributed regime have been
given for an exploding white dwarf in Fig. 2 of \citet{Nie97} and
those conditions define the applicability of this paper.  For those
conditions, $L_{\rm Gib}$ is also much greater than the Kolmogorov
length, $L_{\rm Kol} = L Re^{-3/4} \sim 10^{-4} - 10^{-5}$ cm, where
the turbulence is dissipated. Here $L$ is the integral length scale of
turbulence ($\sim 10^6$ cm) and $Re$ is the Reynolds number
($\sim10^{14}$). The inequality $L_{\rm Kol} \ll L_{\rm Gib} \ll
\delta_{\rm lam}$ is thus satisfied, where $\delta_{\rm lam}$ is the
thickness of the laminar flame (typically $\sim$cm).

Deep in the distributed regime, turbulence is more effective at
transporting both heat and composition, even on scales as small as the
laminar flame width, than conduction and diffusion \citep[the
conductive and radiative opacities here are comparable;][]{Tim00a}. As
a consequence, the concept of a laminar flame, one whose width and
speed are determined by the near equality of burning and diffusion
time scales \citep{Lan59,Tim92}, breaks down. Fuel and hot ash are
co-mingled and the definition of the width of the must be modified
\citep{Lis00a}. If that width grows large enough and the burning
rapid enough, a transition to detonation can occur
\citep{Kho97,Nie97,Lis00b}. Such a transition is impossible within the
extent of a laminar flame; it can only occur in a turbulently stirred
one.

Here the physics of that transition is explored.  Kolmogorov scaling
is assumed throughout. Expressions are derived for the turbulent flame
speed and its width (\Sect{distrib}). A hypothetical ``steady state''
is posited, in which fuel burns at a rate balancing turbulent
mixing. The width of the flame grows as the density declines because
the temperature of the ash, to which the burning rate in the mixture
is very sensitive, is lower there. The turbulent flame moves at a rate
given by the length scale at which turbulent diffusion matches burning
and, as the width of the burning mixture becomes greater, so too does
its speed. Thus, in supernovae, the distributed burning flame moves
faster as the density decreases - the opposite of what happens for a
laminar flame. On larger scales, turbulence also folds these
(turbulently broadened) flames and, just as in the laminar case, there
is an overall ``flame brush'' \citep{Dam40,Hil99} whose motion
determines the total rate of burning.

As the density continues to decline, however, the turbulently mixed
flame grows ever broader, eventually approaching the size of the
integral length scale of the turbulence. Qualitatively, this is the
largest scale at which the anisotropic shear and instabilities
introduced by floatation produce an nearly isotropic cascade of
turbulence with constant energy density. Technically, it is the
distance scale beyond which the self correlation of the velocity
components vanishes. For a typical Type Ia explosion, plumes of size
$\sim$100 km float at speeds in excess of 1000 km
s$^{-1}$. Empirically, from numerical simulations, the size of the
isotropically stirred region is $\sim$10 km, though certainly
variations of a factor of several around this are allowed. Also
important is the velocity at the integral length scale, $u_L$, which
specifies the energy density in the turbulence. This is typically some
fraction, $\sim$10\%, of the floatation speed, but again, large
variations are expected, up to the floatation speed itself.

When the width of the turbulently mixed flame becomes comparable to
the integral length scale of the turbulence itself, the largest
turbulent speeds have thus become an appreciable fraction of the sound
speed. Recent studies by \citet{Roe07a} show the highest turbulent
speeds can approach 1000 km s$^{-1}$ at densities $\sim$10$^7$ g
cm$^{-3}$; the sound speed there is about 4000 - 5000 km s$^{-1}$.  So
long as the assumed steady state solution persists, supersonic burning
remains, by definition, impossible for subsonic turbulence. However,
once there is only a single flame or two in the integral length scale,
the steady state assumption certainly breaks down.  A large eddy and
its accompanying cascade stirs the mixture, but then burning goes on
at a non-steady rate before another eddy happens in the same
region. In \Sect{ddt} and \Sect{ddtcond}, it is shown that this sets
the stage for a detonation. The conditions are restrictive and require
turbulent energies corresponding to close to 1000 km s$^{-1}$ on a
length scale of 10 km {\sl and} a density between 0.5 and $1.5 \times
10^7$ g cm$^{-3}$. This is a much more restrictive condition than just
``entering the distributed burning regime''.

This work bears some similarity to those of \citet{Kho97},
\citet{Nie97}, and \citet{Lis00b}, but considers more carefully both
the conditions in the mixed region and the need for a turbulent flame
that is already moving at a fraction of the sound speed before a
transition to detonation can occur. The conditions required are thus
more precisely determined and much more constrained. The necessity of
a separation between the carbon burning flame and the oxygen burning
flame and the breaking of the steady state assumption at the integral
length scale are emphasized. Some of the conclusions are similar to
those of \citet{Ker01} for low Prandtl number flames, but are more
extensively discussed in the astrophysical context.

\section{A SIMPLE MODEL FOR BURNING IN THE DISTRIBUTED REGIME}
\lSect{distrib}

The basic parameters of any flame, its width and speed, can be
estimated from the rates for fuel consumption and fuel advection into
the burning region.  Although turbulence is stochastic and the
distribution of heat in a stirred region is not nearly so smooth as
when huge numbers of electrons and photons participate in conduction
and diffusion, there is still, on the average, something like a
steady state. In that steady state, the rate at which fuel (carbon) is
brought into the burning region by turbulent eddies balances the rate
of consumption within that region. The whole mixed-up, burning
ensemble moves through the fuel with a typical speed which is
essentially the size of the region divided by its turbulent turnover
time.  That is,
\begin{equation}
\int \frac{dn_{12}}{dt} dV = \int n_{12} v_T dA,
\end{equation}
where $n_{12} = \rho N_{A} Y_{12}$ is the number density of carbon
nuclei as a function of location, $\rho$ is the density, $N_{A}$,
Avogadro's number, $Y_{12}$, the mass fraction of carbon, $X_{12}$,
divided by 12, $v_T$, the velocity of the burning turbulent front
normal to the area, $A$, and $V$, the volume bounded by that area.
Here we consider a one-dimensional flame in plane geometry. For the
low densities of interest, reactions beyond carbon burning, i.e.,
oxygen burning, occur so far behind the carbon burning flame as to be
negligible on the scale of the problem.  Let the thickness of the
mixed region be $\lambda$, then
\begin{equation}
\int_{0}^\lambda \rho(l') \frac{d Y_{12}(l')}{dt} \, d l' \ = \ \rho_{\rm fuel} Y_{12}^0 v_T
\lEq{steady}
\end {equation}

The left hand side gives the rate of carbon destruction (in gm
cm$^{-2}$ s$^{-1}$) by nuclear reactions, while the right hand side is
the rate of advection into the burning region by turbulent
eddies with characteristic length scale, $\lambda$, and speed
$v_T$. Here, $\rho_{\rm fuel}$ is the density in the unmixed fuel and,
$Y_{12}^0$, the carbon abundance there. The burning rate is
\begin{equation}
\frac{d Y_{12}}{dt} \ = \ -2 \, \rho Y_{12}^2 \, R_{12,12}(\rho,T),
\end{equation}
where $R_{12,12}$ is the rate factor for the carbon fusion reaction.
Because of the temperature sensitivity of this rate, about $T^{20}$,
most of the carbon consumption goes on in a narrow region where the
mass fraction is low and the temperature high \citep{Bel04}, closer to
the hot ash than to the cold fuel. The rapid rate of burning there has
to balance, on the average, what is advected into the larger
region. In order to obtain $v_T$, it is thus necessary to specify
$\rho(l')$, $T(l')$, $Y_{12}(l')$, and $v_T$.

For $v_T$, it is appropriate to take the velocity of the turbulent
eddy with length scale $\lambda$.  Assuming Kolmogorov scaling and a
turbulent energy input on the integral length scale, L, corresponding
to velocity $u_L$,
\begin{equation}
v_T \ = \ \left(\frac{\lambda}{L}\right)^{1/3} u_L
\lEq{kolm}
\end {equation}
where, from typical numerical simulations, $u_L$ is in the range
10$^7$ - 10$^8$ cm s$^{-1}$ for L = $10^6$ cm \citep{Roe07a}.

All speeds are very subsonic, so to good approximation, the pressure
in the fuel, ash, and the mixture is constant.  Provided that mixing
is faster than burning and conduction, the temperature and density in
the mixture are thus uniquely defined by the fractions of ash and fuel
that are mixed and the compositions of each. In fact, some burning
does occur during the mixing and this affects the temperature
distribution, but to first order, the temperature obtained by burning
to a certain carbon mass fraction is the same as that obtained by
mixing cold fuel with ash of higher energy and lower carbon abundance
to obtain that same final mass fraction. This is not precise because
carbon burns to different products at different temperatures and so
the energy released is not a linear function of carbon consumed, but
the difference is not large.

Next we seek a description of how temperature and density vary in the
mixed region. This requires an {\sl ansatz} for how the carbon
mass fraction varies.  To illustrate the procedure, assume an initial
composition of 50\% C and 50\% O at a density of $1.0 \times 10^7$ g
cm$^{-3}$ and temperature $6.0 \times 10^8$ K. This is a typical
temperature in the outer parts of the white dwarf when it runs away,
but the answer will not depend on the exact value because the pressure
is not very sensitive to the temperature and the internal energy of
the ash is much higher than that of the fuel.  The pressure in this
fuel is $9.046 \times 10^{23}$ dyne cm$^{-2}$ and its internal energy
is $\epsilon = 1.706 \times 10^{17}$ erg g$^{-1}$ \citep[here and
throughout the paper we employ the Helmholtz equation-of-state routine
of][]{Tim00b}. Now mix this fuel with a small amount of ash so that
the temperature rises a small increment, $\delta T$, and the carbon
fraction goes down. For this new state, (T,P), iterate on the density
and internal energy until a solution is found with the same pressure
as before. The new density is lower and its formation required
expansion. Its new energy is the heat brought in by the mixing minus
the energy lost to PdV work. That is
\begin{eqnarray}
\delta q \ &= \ \epsilon(T_o+\delta T) - \epsilon(T_o) \ + \ 
P \ \left(\frac{1}{\rho +\delta \rho} - \frac{1}{\rho}\right) \\
\nonumber
&\approx \ \epsilon(T_o+\delta T) - \epsilon(T_o) \ - \ 
P (\frac{\delta \rho}{\rho^2}).
\end{eqnarray}
Here $\delta \rho$ is inherently negative so the pressure term is
positive.  The composition of the mixture will also have changed by an
amount that depends upon the composition of the ash.  Here we adopt
the ash composition given by following isobaric burning to completion
off-line using a small 7 isotope network.  For the conditions given
above, a typical ash composition will be 57\% O, 16\% Mg, 26\% Si and
1\% S. At a density of $3 \times 10^7$ g cm$^{-3}$, the composition
would have been slightly different: 57\% O, 8\% Mg, 33\% Si, and 2\%
S, but the Q-value is not very sensitive to the difference. The change
in nuclear binding energy between the ash and fuel (50\% each $^{12}$C
and $^{16}$O) is thus $Q = 3.10 \times 10^{17}$ erg
g$^{-1}$. Substantially different numbers characterize an initially
carbon-rich composition. For a fuel that is 75\% carbon and 25\%
oxygen, the ash composition at 10$^7$ g cm$^{-3}$ is 38\% O, 15\% Mg,
45\% Si, and 2\% S, implying Q = $4.70 \times 10^{17}$ erg g$^{-1}$.

The fraction of carbon in the mixture is
\begin{equation}
Y_{12}(T+\delta T) = Y_{12}({\rm fuel}) (1 \ - \ \frac{\delta q}{Q}),
\end{equation}
since, by definition, $Y_{12}$(ash) = 0. The other composition
variables are similarly interpolated between their initial (fuel) and
final (ash) values based upon the change in energy.  Using this new
composition, the density is again iterated to find the isobaric state
appropriate to the new temperature and self-consistent mixed
composition. The process is continued for about 1000 steps until the
carbon abundance is zero. The outcome of this calculation is a set of
temperatures, densities, and compositions for the mixture consistent
with the pressure in the fuel.

To reach closure, it remains to specify how the carbon abundance
varies within the mixed region. In reality, the carbon mass fraction
will be heterogeneous, reflecting the operation of numerous
eddies on all scales and the large Lewis number. Stirring is more
effective on small scales so the most natural distribution would be a
``noisy'' staircase function, or even a homogeneous mixture
\citep{Ker01}. We return to this picture in \Sect{ddt}.  For
now, however, a simple approximation is made that the carbon abundance
is distributed linearly within the stirred flame.  That is, for $0 \le
l'/\lambda \le 1$,
\begin{equation}
Y_{12} \ = \ Y_{12}^o \ (1 - \frac{l'}{\lambda}).
\end{equation}
Such a linear approximation is consistent with multi-dimensional
simulations so far \citep[Fig. 31 of ][]{Bel04} and gives equations that are
easy to manipulate and understand.

Given $Y_{12}(l')$, $\rho(Y_{12})$ and $T(Y_{12})$, L, and $u_L$, one
is now equipped to solve \Eq{steady} for a unique value of $\lambda$.
Some results are given in \Fig{flame} and Table 1, the latter showing
a dramatic dependence of the burning front width and speed on the
density and turbulent energy.  For the lowest densities and highest
turbulent energies considered, the mixed region becomes comparable to
the integral length scale, 10 km, and the burning can approach a
fraction of the sound speed. The numbers which give $\lambda \gtaprx$
10 km in Table 1 are not physical unless eq. (5) and the assumption of
homogeneous, isotropic turbulence can be extrapolated to these larger
scales, and the velocity continues to increase above $u_L$. In those
cases where $u_L$ is already $10^8$ cm s$^{-1}$, that is doubtful.
The flame widths and speeds are also smaller for carbon-rich mixtures.
Burning more carbon raises the temperature of the ash and mixture and
makes the burning region smaller.

It is important that the flame has separated into two components.  If
one added the energy generation from oxygen burning, the temperatures
would be higher and the width of the burning region much smaller.
Since as we shall see, only the largest values for $\lambda$ in Table 1
imply a possible transition to detonation, a necessary condition for a
delayed detonation transition is that the oxygen and carbon burning
flames have split and are widely separated.

\section{Approximations to the Turbulent Flame Speed}
\lSect{approx}

The speed of a laminar flame is proportional to the
square root of the heat diffusion coefficient. In the distributed
regime one expects a similar relation with the turbulent diffusion
coefficient, $D_{\rm turb} \sim v_{\lambda} \lambda$, substituting for
the radiative one \citep[e.g.,][]{Roe05}. Here $v_{\lambda}$ is the turbulent
eddy speed on the scale of the flame width, $\lambda$, and hence
$D_{\rm turb} \sim \lambda^{4/3} u_L L^{-1/3}$. The width of the flame 
is given by equating the nuclear and diffusion times
\begin{equation}
\tau_{\rm nuc} \approx \frac{\lambda^2}{D_{\rm turb}} = \frac{L^{1/3}\lambda^{2/3}}{u_L},
\end{equation}
and hence
\begin{equation}
\lambda \approx \frac{(\tau_{\rm nuc} u_L)^{3/2}}{L^{1/2}}.
\lEq{lambda}
\end{equation}
This relation is well known in the chemical combustion community
\citep{Pet99,Pet00,Ker01}.  Here $\tau_{\rm nuc}$ is the average
nuclear burning time in the region defined by eq. (2). For a given
density and turbulent energy, this suggests a scaling $\lambda \propto
u_L^{3/2}$ which agrees with the values in Table 1. Such a scaling is
also expected because the left hand side of \Eq{steady} is
proportional to $\lambda$ while the right hand side depends on
$\lambda^{1/3} u_L$. \EQ{lambda} also states that the turbulent flame
speed, $\lambda/\tau_{\rm nuc}$, is the square root of the energy
dissipated by the turbulent cascade, $u_L^3/L$ in a nuclear time
scale. For a given turbulent energy, to get the flame to move faster
one must increase the nuclear time scale, i.e., slow the burning.

The nuclear time scale is very sensitive to the temperature in the
mixed region which is highly variable, but near $3 \times 10^9$ K, it
is approximately \citep{Woo04}
\begin{equation} 
\tau_{\rm nuc} \approx  \frac{C_P T}{n \dot S_{\rm nuc}},
\end{equation}
where $\dot S_{\rm nuc} \propto \rho X_{12}^2 T^n$. The heat capacity, due
to a combination of semi-degenerate electrons and radiation, increases
{very} roughly as $T^2$ while, for barrier penetration, $n$ in the
temperature range near 2 to 3 billion K is
\begin{equation}
n = 28.05 \, T_9^{-1/3}  -  \frac{2}{3} \approx 20.
\end{equation} 
Thus $\tau_{\rm nuc}$ is roughly proportional to $\rho^{-1}
X_{12}^{-2} T^{-17}$ and, since the temperature in the burning region
is proportional to $T_{\rm ash}$, the flame width,
\begin{eqnarray}
\lambda \ \propto& \ u_L^{3/2} L^{-1/2} \rho_{\rm fuel}^{-3/2}
T_{\rm ash}^{-25.5} X_{12}^{-3} \\ 
\nonumber
\approx& \ 4.9 \
(\frac{T_{\rm ash}}{2.79 \times 10^9 \ {\rm K}})^{-25.5}
(\frac{u_L}{10^8 \ {\rm cm \ s^{-1}}})^{3/2} \\
\nonumber
&\ \ \ \ \ \ \ \ (\frac{10 \ {\rm km}}{L})^{1/2} 
(\frac{\rho_{\rm fuel}}{10^7 \ {\rm g
\ cm^{-3}}})^{-3/2} (\frac{0.5}{X_{12}})^3 \ \ \ {\rm km},
\lEq{width}
\end{eqnarray}
where the proportionality has been normalized to the numerical results
in Table 1 for the fiducial values. This is an overall good fit
for other densities and compositions in the range of interest.

The turbulent speed, $v_T$, is given either by \Eq{kolm} evaluated for
the flame width derived above, or by
\begin{eqnarray}
v_T  \approx& \ \frac{\lambda}{\tau_{nuc}} \\
\nonumber
\approx& \ \frac{u_L^{3/2} \tau_{\rm nuc}^{1/2}}{L^{1/2}} \\
\nonumber
\approx& \  \ 790 \ (\frac{T_{\rm ash}}{2.79 \times 10^9 \ {\rm K}})^{-8.5}
(\frac{u_L}{10^8 \ {\rm cm \ s^{-1}}})^{3/2} \\
\nonumber
&\ \ \ \ \ \ \ \ (\frac{\rho_{\rm fuel}}{10^7 \ {\rm g \ cm^{-3}}})^{-1/2} 
(\frac{0.5}{X_{12}}) (\frac{10 \ {\rm km}}{L})^{1/2} \ \ \ {\rm km \ s^{-1}}.
\lEq{speed}
\end{eqnarray}
In addition, $v_T$ should not be greater than $u_L$.  Note the strong
reciprocal dependence on the temperature of the ash and hence on the
density and energy yield of the burning. It is this dependence that is
chiefly responsible for the spreading and acceleration of the
turbulent flame at low density. If oxygen burned as well as carbon,
$T_{\rm ash}$ would be much greater and the flame speeds and widths
would be drastically reduced.  The fast speeds and broad widths
derived here rely on the oxygen flame lagging far behind the carbon
flame, i.e., outside the mixed region. For the low densities we
consider, this is the case. Note also that $v_T$ is enormously greater
than the laminar flame speed, which at 10$^7$ g cm$^{-3}$ is only 3000
cm s$^{-1}$ \citep{Bel04}.

The quantity $T_{\rm ash}$ can be computed off line as a function of
initial density and composition \citep{Tim92}. Approximate values
obtained using a small 7 isotope network are given in Table 2.  The
values in the table can be approximately fit by an expression of the
form
\begin{equation}
T_{\rm ash} \ \approx \ 2.79 \times 10^9 \left(\frac{\rho_{\rm fuel}
X_{12}}{5 \times 10^6 \ {\rm g \ cm^{-3}}}\right)^{0.25}. 
\lEq{tash}
\end{equation}
Substituting this in \Eq{width} and \Eq{speed}, one has
\begin{eqnarray}
\lambda \ \approx& \ 4.9 
\ (\frac{u_L}{10^8 \ {\rm cm \ s^{-1}}})^{3/2} (\frac{10 \ {\rm km}}{L})^{1/2}
\\
\nonumber
 &\ \ \ \ \ \ \ \ (\frac{\rho_{\rm fuel}}{10^7 \ {\rm g
    \ cm^{-3}}})^{-7.9} (\frac{0.5}{X_{12}})^{9.4} \ \ \ {\rm km},
\lEq{width1}
\end{eqnarray}
and
\begin{eqnarray}
v_T \ \approx& \  \ 790 \ 
(\frac{u_L}{10^8 \ {\rm cm \ s^{-1}}})^{3/2} \\
\nonumber
&\ \ \ \ \ \ \ \ (\frac{\rho_{\rm fuel}}{10^7 \ {\rm g \ cm^{-3}}})^{-2.6} 
(\frac{0.5}{X_{12}})^{3.1} (\frac{10 \ {\rm km}}{L})^{1/2} \ \ \ {\rm km \ s^{-1}}.
\lEq{speed1}
\end{eqnarray}

For a given turbulent energy and carbon fraction, this implies a flame
speed that scales roughly as $\rho^{-2.6}$

\section{THE TRANSITION FROM A DEFLAGRATION TO A DETONATION}

\subsection{Spontaneous Detonation}
\lSect{spontaneous}

One of the simplest ways a transition to detonation could happen in an
exploding white dwarf, which is included here only because it seems to
have been overlooked, is if frictional heating - the dissipation of
the turbulent energy on the Kolmogorov scale - heats a region of fuel
to the flash point.

Consider a region of fuel close to a large rising element of ash.  The
rise of a burning plume injects turbulent energy at some
characteristic length scale. The velocities are initially anisotropic,
but after the energy cascades down approximately one decade in length
scale, that energy resides in isotropic Kolmogorov turbulence
\citep{Zin05}. Typical turbulent speeds at length scales of $\sim$10
km where the Kolmogorov cascade might begin are, at reasonably late
times in the explosion, in the range 1 - 10 $\times 10^7$ cm s$^{-1}$
\citep{Sch06,Roe07a}.

This energy cascades downwards to the Kolmogorov length, 10$^{-4}$ -
10$^{-5}$ cm, where it dissipates as heat. The amount of heat
dissipated is $u_L^3/L$ erg g s$^{-1}$, i.e., the conserved quantity
in Kolmogorov turbulence. For speeds 10$^7$ - 10$^8$ cm s$^{-1}$ on a
length scale of 10 km, that corresponds to 10$^{15}$ - 10$^{18}$ erg g
s$^{-1}$.  Most of this dissipation occurs inside the ash
\citep{Sch06}, in part because the flame spreads at a speed comparable
to that of largest eddies. However, there may be small regions near
the floating ash, unresolved in current studies, perhaps within the
Kelvin-Helmholtz rolls that bound the rising plumes or in the wakes of
detached bubbles, where a locally large concentration of turbulent
energy is dissipated in the fuel. The dissipation might be
particularly large in vortex tubes shed by the rising plumes.

Because of its low heat capacity, $\ltaprx 2 \times 10^{17}$ erg
g$^{-1}$ is necessary to raise the fuel temperature to the point where
it will burn supersonically. If the explosive burning region is larger
than a critical mass, a detonation will occur \citep{Nie97,Dur06}. For
the highest turbulent energies considered, this would only take about
0.2 s of uninterrupted dissipation, significantly less than the
expansion time of the star. For regions in close proximity to an
active flame, the frictionally heated fuel will probably be burned
before it can run away. However, in the trailing wake of rising
bubbles, there might be time for viscous dissipation in unburned fuel
to ignite new burning. If the temperature gradient in this new region
is sufficiently small, the ignition could have a supersonic phase
velocity.

\subsection{Detonation in Fuel-Ash Mixtures}
\lSect{ddt}

In the absence of this viscous ignition (\Sect{spontaneous}), or
ignition by compression \citep{Arn94,Ple04,Roe07b}, carbon detonation
can only occur in a mixture of hot ash and cold fuel in the
distributed regime.  But there can be no detonation within the mixture
so long as a {\sl steady state subsonic} flame exists, with burning
balancing the average rate at which fuel is heated either by
conduction or mixing. This is always true in the case of laminar
flames where the thickness is much less than the critical mass for
detonation, but it remains true for steady state flames in the
distributed regime. Even though the widths of the flames in Table 1
can become quite large, the turbulent eddy that sets the time scale
for burning in $\lambda$ is itself subsonic. Burning occurs at a rate
just sufficient to balance the advancement of the mixing and cannot
become supersonic.

This steady state is a fiction though, useful only for obtaining rough
estimates for the size and speed of the mixed burning region. So long
as the mixing region defined by the turbulent integral length scale
contains many flames, fluctuations in the burning rate will average
out.  The situation remains closely analogous to the flamelet
regime. Many flame surfaces combine to make a ``flame brush'' with
fractal dimension D = 2.36. For the allowed range of length scales,
the burning region moves at a speed given by the largest turbulent
eddies and the individual flame speeds, be they turbulent or laminar,
are not important.

The situation changes though when the entire integral length scale of
the turbulence contains only one or a few flames (Table 1) The large
eddies driving the mixing are random. Occasionally, a long time may
elapse before a new eddy arrives. Within the region stirred by this
large eddy, layers exist of nearly isothermal mixtures of fuel and
ash. Such layers were seen by \citep{Lis00a} and were responsible for
the ``micro-explosions'' observed in their simulations, but because of the
small dimensions of the mixed flames studied in that paper by Lisewski
et al., that burning never approached sonic speeds.

\subsection{Conditions for Detonation}
\lSect{ddtcond}

A necessary condition for detonation is that sustained burning inside a
distributed flame width occur faster than its sound crossing
time. This is a condition on the sonic length scale,
\begin{equation}
r_{\rm sonic} \ =  \ c_s \ \tau_{\rm nuc}(T') \ \ltaprx \ \lambda \ < \ L. 
\end{equation}
where $T'$ is some temperature $0 < T < T_{\rm ash}$ in the isobaric
mixture. If $T'$ in this equation were the same as the temperature in
$\tau_{\rm nuc}$ in \Eq{lambda}, one would require supersonic
turbulent motion in order to initiate a detonation, since that would
imply
\begin{equation}
L \ > \ \lambda \ \gtaprx \ \left(\frac{c_s^3 }{u_L^3}\right) L.
\end{equation}

The fallacy in this argument is that $\lambda$ is some approximate
length scale in a fictitious steady state, which never exists at any
one place and time, while $r_{\rm sonic}$ can vary greatly depending
on the instantaneous local values in a given flame.  Occasionally the
distribution of temperature inside the mixed flame is such that,
including the effects of induction, it burns much faster than steady
state. Table 3, calculated using a small 7-isotope network, gives the
characteristics of burning on different time scales. Assume, as we
shall find is necessary, that a high degree of turbulence exists such
that $L/u_L \sim 0.01$ s (i.e., $u_L \sim 10^8$ cm s$^{-1}$, $L \sim
10^6$ cm). Mixing can go on without appreciable burning so long as the
temperature, T, remains cool enough that $\tau_{\rm nuc}(T) \gtaprx
L/u_L$. Here a small margin of error is included, and the temperature
is calculated such that half the fuel would burn in 0.02 s. This is
$T_{0.02}$ in Table 3 and the corresponding carbon mass fraction is
X$_{12,0.02}$. The mixture and time scale are calculated for isobaric
fuel-ash mixtures with an initial carbon mass fraction of either 0.5
or 0.75. Because carbon burns to different compositions at different
temperatures, this does not give exactly the same values as burning a
given composition in place without mixing, but for the mixing that
goes on before burning this is the correct procedure.

These mixed plasmas are then allowed to burn without further
intervention.  Because of the high temperature sensitivity of the
reaction rate, most of the energy from burning is released after
mixing during this phase of ``inductive burning''. Towards the end,
the burning can becomes supersonic for a region, $r_{\rm sonic}$,
given by the sound speed and burning time.  The minimal burning time,
hence smallest, $r_{\rm sonic}$, called here $r_{\rm sonic}^{\rm
  min}$, is evaluated by two criteria. One is that 60\% of the initial
carbon has burned. This corresponds to the onset of the rapid rise in
e.g., \Fig{flame}. A second value, the actual minimum value of $r_{\rm
  sonic}$ comes from numerically evaluating the point where the energy
generation is a maximum for an isobaric mixture of carbon and ash
starting at the initial temperature, density and carbon mass
fraction. The initial temperature matters little. The first choice
gives a lower temperature and hence larger $r_{\rm sonic}$ and is thus
more difficult to achieve. However, it gives a larger carbon mass
fraction which decreases the necessary mass for detonation.  This first
set of numbers was plotted for X$_{12,{\rm initial}}$ = 0.5 in Fig. 4.

For fuel densities below $7 \times 10^6$ g cm$^{-3}$ and $X_{12}$ =
0.50, the ash temperature is just too low to burn in less than
0.02 s, no matter what the mixture. The sonic radius is much
larger than the integral length scale and starting to be a significant
fraction of the white dwarf itself. It seems that spontaneous
detonation will be impossible below this density and a carbon mass
fraction of 0.50. Somewhat lower densities can be accommodated in the
carbon abundance is 0.75, but no lower than $5 \times 10^6$ g
cm$^{-3}$.

The sonic radii in Table 3 can be fit by a function
\begin{equation}
r_{\rm sonic}^{\rm min} \ \approx \ 3.5 \ {\rm km} \ F \ \rho_{\rm
  7,fuel}^{-4.5} \left(\frac{0.5}{X_{12}}\right)^6,
\lEq{sonicr}
\end{equation}
where $0.4 \ltaprx F \ltaprx 1$. An alternate derivation that assumes
a constant sound speed (4000 - 5000 km s$^{-1}$) and evaluates the
nuclear time scale assuming $T_{\rm max} \propto T_{\rm ash}$ gives a
similar result.

The condition $r_{\rm sonic}^{min} \ltaprx \lambda$ in \Eq{width1} gives
\begin{eqnarray}
\rho_{\rm DDT} \ \ltaprx& \ 1.1 \times 10^7 \ {\rm g \ cm^{-3}}
\left(\frac{0.5}{X_{12}}\right)^{0.70} \\ 
\nonumber
& \ \ \ \ \ \left(\frac{u_L}{10^8 \ {\rm cm \ s^{-1}}}\right)^{0.55}
F^{-0.29} \ \left(\frac{10 \ {\rm km}}{L}\right)^{0.15}.
\lEq{ddtcabun}
\end{eqnarray}
In addition, the condition that $r_{\rm sonic}^{\rm min}$ be less than L
$\sim 10$ km gives 
\begin{equation}
\rho_{\rm DDT} \ \gtaprx \ 8 \times 10^6 \ {\rm g
  \ cm^{-3}} \ F^{0.22} \left(\frac{0.5}{X_{12}}\right)^{1.3} L_{10}^{-0.2}
\end{equation}

To check whether the conditions derived here are not only necessary
but sufficient, an offline calculation of detonation was carried out
using the Kepler code \citep{Wea78,Woo02} and the procedure defined in
\citet{Nie97}. Arguably, this might err on the conservative side since
it assumes a spherical distribution and the shock wave suffers some
geometric dilution moving out, but the procedure has been shown to
give qualitative agreement with more rigorous calculations
\citep{Dur06}. A region of size of variable length, $L_1$, with
characteristics given in Table 3 ($T_{\rm max}, \rho_{\rm max}$) was
embedded in a comparable size region, $L_2$ where the temperature
declined gradually to the cold fuel value. A sample set up was shown
in \Fig{rsoninit}. It is important that the isothermal region is
surrounded by a region where the temperature changes gradually so that
a supersonic phase velocity can develop.  This gradual change in
temperature seems reasonable given the turbulent mixing that occurs on
all length scales.

The results for some trial detonation are given in Table 4 and
\Fig{rsondet}. It seems that homogeneously mixed regions of size
greater than a few times $r_{\rm sonic}^{min}$ are capable of igniting
detonations. The condition $r_{\rm sonic}^{\rm min} << \lambda$ is not
only necessary for detonation, but sufficient.  The fact that the size
is somewhat larger than $r_{\rm sonic}$ is not alarming because
ignition may involve several adjacent mixed layers.

\subsection{Dependence on the carbon abundance}
\lSect{cabun}

The carbon abundance affects the transition in several ways, sometimes
subtly. One might expect that, due to its larger energy release,
burning a more ``incendiary'' mixture of carbon and oxygen, one with a
larger carbon mass fraction, would make it somehow more likely to
detonate. In fact, \Eq{ddtcabun} shows the opposite behavior, a weak
reciprocal dependence of the detonation density upon the local carbon
abundance.  This is because the burning of a carbon-rich composition
produces a hotter ash, and the fuel ash mixture maintains a thin
burning width until a lower density. This is unavoidable.

Unfortunately, this is the opposite behavior to that postulated by
\citet{Ume99} in an attempt to explain the preponderance of bright
Type Ia supernovae in late type galaxies \citep{Fil89,Bra96}. The
transition densities computed here are also lower and the dependence
on density, comparatively weak. 

However, there is another possibility. \EQ{ddtcabun} is a necessary,
but not sufficient condition for detonation. To detonate, one also
needs a critical mass. Table 4 shows that when the homogeneously mixed
region (Table 3) is two to three times $r_{\rm sonic}^{min}$
detonation usually happens. Only one model is this paper strictly
satisfies the criteria $L_1 \sim 3 r_{\rm sonic}^{\rm min} \ltaprx
\lambda$ for $u_L \ltaprx 10^8$ cm s$^{-1}$ at 10 km and that is the
model with X$_{12}$ = 0.75 and $\rho_{\rm fuel}$ = $6 \times 10^6$ g
cm$^{-3}$. Turning the density down to find a solution where this
works for $X_{12}$ = 0.5 requires mixed regions that are either
considerably larger than the integral length scale, or so cold that
they do not detonate at all.  Certainly the accuracy of this study is
not enough to rule out models with $X_{12}$ = 0.5, which do satisfy this
condition within a factor of about 2 ($X_{12}$ = 0.5, $\rho_{rm fuel}
= 10^7$ g cm$^{-3}$, for example, but a possibility is that detonation
requires a certain minimum value of carbon mass fraction. This would
have major ramifications. Models with less than critical $X_{12}$
would have to be pure deflagrations and would have distinctly
different properties. A lot more study is needed before this
possibility is taken too seriously.

Alternatively, there may be other explanations for the preponderance
of bright Type Ia supernova in late type galaxies. Perhaps, the
carbon-oxygen ratio in the outer layers of the white dwarf at the time
it explodes correlates differently with stellar evolution than Umeda
et al. assume. In this regard it is noteworthy that more massive white
dwarfs, which are derived from larger stars, {\sl do} have lower
carbon abundances in their interiors when they explode \citep[Fig. 6
  and 12 of][]{Ume99b}. The carbon abundance, and other factors, may
also influence the ignition conditions which are known to have a major
effect on the supernova brightness.

\section{CONCLUSIONS}
\lSect{conclu}

Using simple scaling arguments, the steady state width and speed of a
turbulent flame in the distributed regime have been derived as a
function of turbulent energy, density, and carbon mass fraction. This
speed is much faster than the laminar speed at the same density, and
displays a sensitive reciprocal dependence on the density. All such
steady state flames are subsonic so long as the turbulence driving
them remains subsonic. The bulk propagation of the burning in the
distributed regime, prior to any detonation, will still be governed by
the motion of the large turbulent eddies, not the speed of the
individual flames.

As the density nears $1 \times 10^7$ g cm$^{-3}$, however, the width
of the flame approaches the integral length scale and the unsteady
nature of the burning becomes important. For sufficiently intense
turbulence, detonation becomes possible \citep[see also][]{Lis00b}.
In order for detonation to occur, several criteria must be satisfied
simultaneously.

First, the carbon and oxygen burning flames must separate
spatially. If both fuels burn simultaneously, the resulting ash is too
hot, and the fuel-ash mixture, too combustible. This makes the average
width of the mixed burning region narrow and prevents the detonation
due to flame broadening described here.

Second, the speed of the largest turbulent eddies must approach the
sound speed. That is, the maximum speed below which the assumption of
isotropic Kolmogorov turbulence is valid must not be too subsonic.  It
is this speed that sets the characteristic scale of the problem. Some
additional increase in flame speed may achieved by unsteady burning,
but the amplification is not likely to be large enough to bridge
orders of magnitude. It is the large turbulent energies reported by
\citet{Roe07a}, $\sim 20$\% sonic at the integral length scale, that
makes things work.

Third, the size of the mixture of fuel and ash must become as large as
the largest eddies. That is, the flame width must approach the
integral length scale. For white dwarfs this is some multiple of 10
km. The most likely detonation site will be at the merger of several
such flames.

These conditions may occasionally all be satisfied at the low
densities encountered by the flame as it moves to the surface layers
and the white dwarf begins to come apart. However, just reaching a
particular density is not adequate. Nor is it sufficient simply to
move into a region where the burning is distributed \citep[e.g., the
  Gibson length is smaller than the flame thickness][]{Roe07c}. Most
of the models in Table 1 are in the distributed regime, but only those
where the flame width $\lambda$ approaches the integral length scale,
L, can detonate.  Detonation thus requires the right combination of
turbulent energy and density and such conditions are rare. Whether
they are sufficiently rare as sometimes not to happen is beyond the
scope of this paper.

It is interesting though that the range of detonation densities
derived here, $0.5 - 1.5 \times 10^7$ g cm$^{-3}$, is what has been
invoked for some time in order to achieve good agreement with
nucleosynthesis, spectra, and light curves in artificially
parametrized descriptions of the explosion \citep{Hoe95}.

Many of the approximations in this paper and its conclusions warrant
careful checking by numerical experiments that this paper will
hopefully motivate.

\acknowledgements

The author gratefully acknowledges helpful conversations on the
subject of the paper with John Bell, Martin Lisewski, Jens Niemeyer,
Fritz Roepke, Mike Zingale, and especially Alan Kerstein.  This
research has been supported by the NASA Theory Program NNG05GG08G and
the DOE SciDAC Program (DE-FC02-06ER41438).

\begin{figure}
\includegraphics[angle=90,width=\columnwidth]{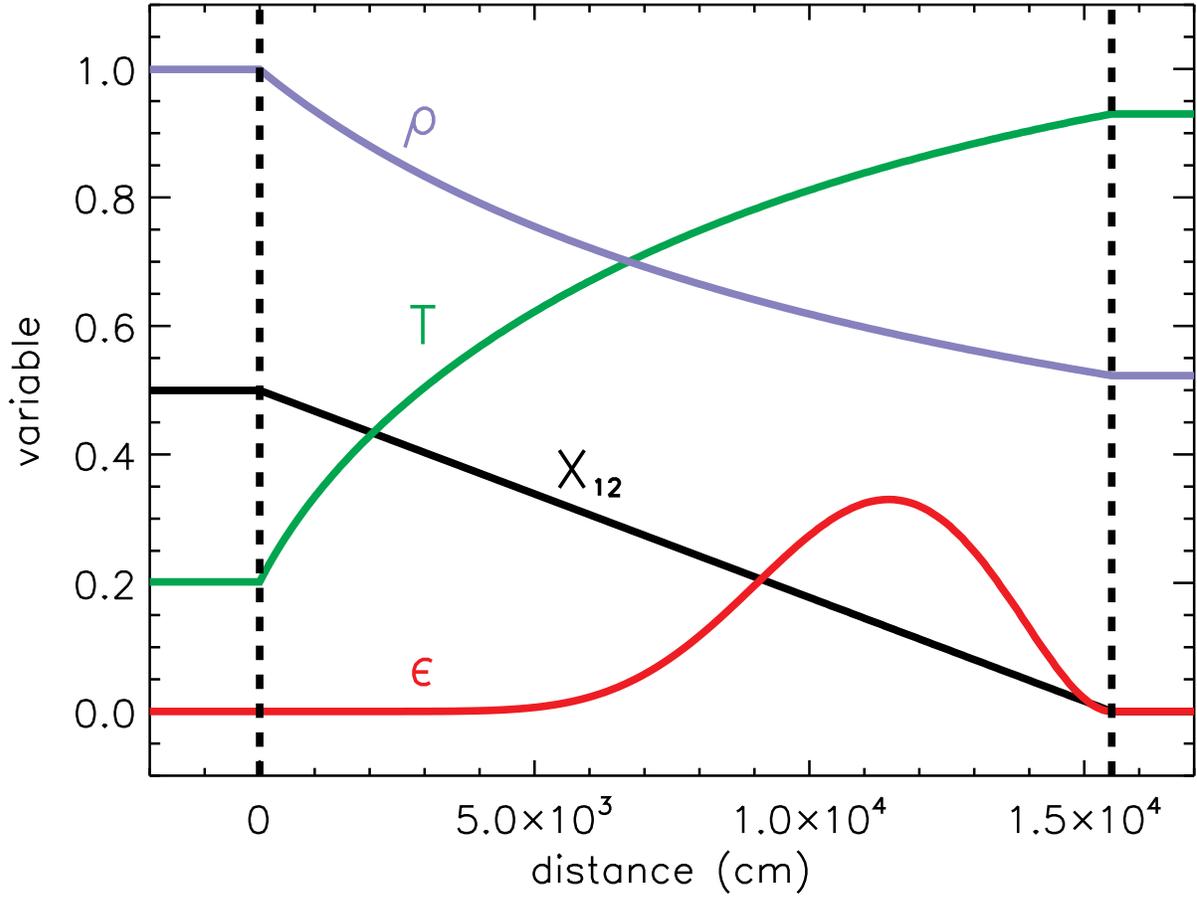}
\caption{Variables in an idealized steady state flame in the
  distributed regime for a density $1 \times 10^7$ g cm$^{-3}$ and
  turbulent speed 10$^7$ cm s$^{-1}$ on a length scale of 10 km. For
  other turbulent energies, the x-axis would be multiplied by
  $(u_L/10^7)^{3/2}$ where $u_L$ is the turbulent speed on the
  integral length scale (10$^6$ cm) in cm s$^{-1}$. Variables have
  been renormalized for plotting. The temperature has been divided by
  $3 \times 10^9$ K; the density, by $1.0 \times 10^7$ g cm$^{-3}$,
  and the carbon consumption rate, $\epsilon$, by 100 s$^{-1}$. To the
  left of the vertical dashed line at the origin is unburned fuel. To
  the right of the vertical dashed line at 0.154 km is ash. For the
  assumed turbulent speeds, the whole distribution is moving to the
  left with a steady state speed of 25 km s$^{-1}$ (see Table
  1). \lFig{flame}}
\end{figure}

\begin{figure}
\includegraphics[angle=90,width=\columnwidth]{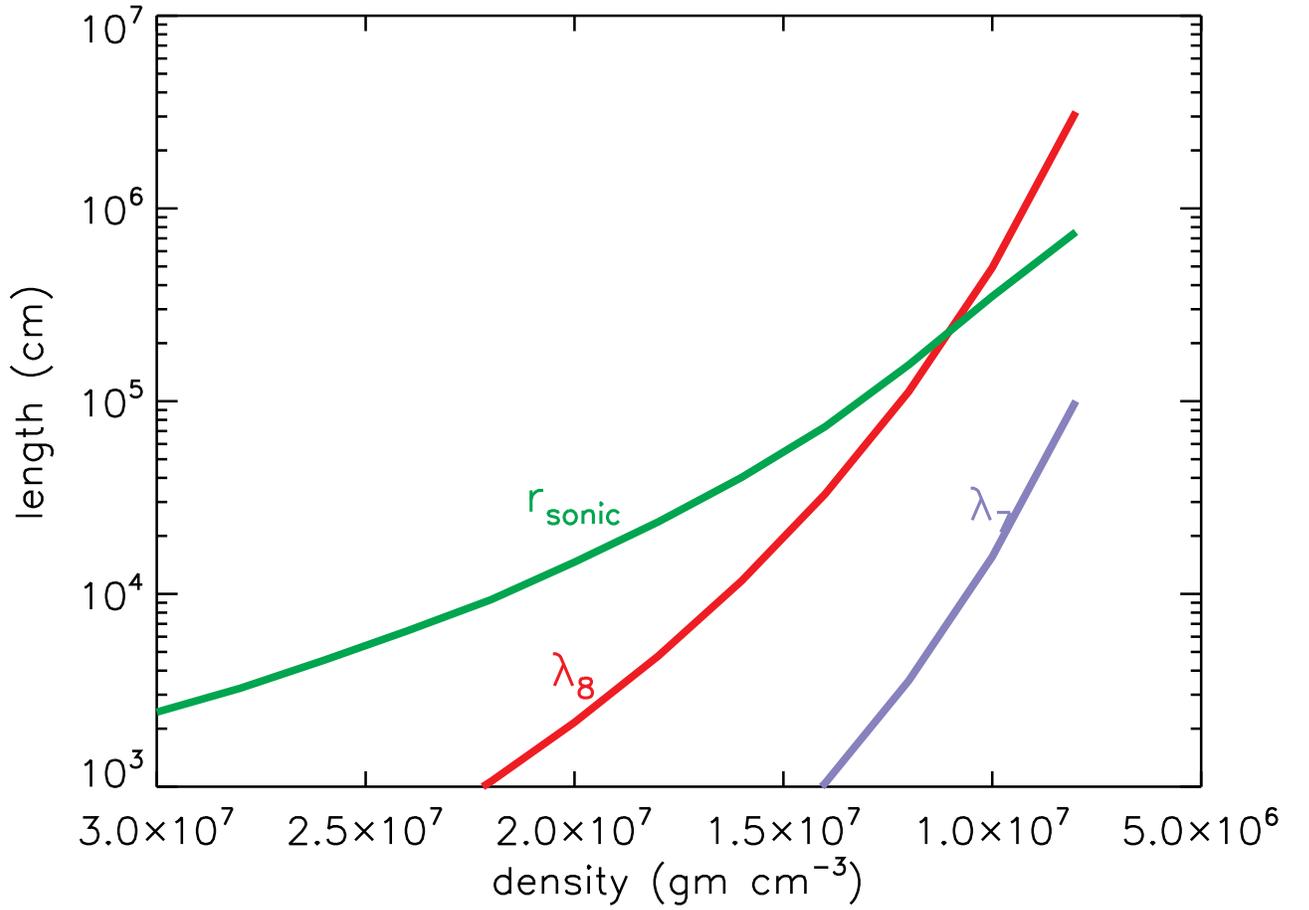}
\caption{For a carbon mass fraction in the fuel of 50\%, the figure
  shows the relative sizes of the sonic radius (see text) and the
  turbulent flame widths for two values of turbulent energy
  characterized by velocities on a scale of 10 km of 10$^7$ cm
  s$^{-1}$ ($\lambda_7$) and 10$^8$ cm s$^{-1}$ ($\lambda_8$) - see
  also Table 1. For densities much greater than 10$^7$ g cm$^{-3}$,
  despite being in the distributed burning regime, the flame will be
  too thin to contain a region capable of running away
  supersonically. For lower density, however, and large turbulent
  energies, detonation is possible. No solutions smaller than 100 km
  could be found for $r_{\rm sonic}$ for densities below $7 \times
  10^6$ g cm$^{-3}$. For the smaller turbulent energy shown, detonation
  is unlikely. \lFig{rsonic}}
\end{figure}

\begin{figure}
\includegraphics[angle=90,width=\columnwidth]{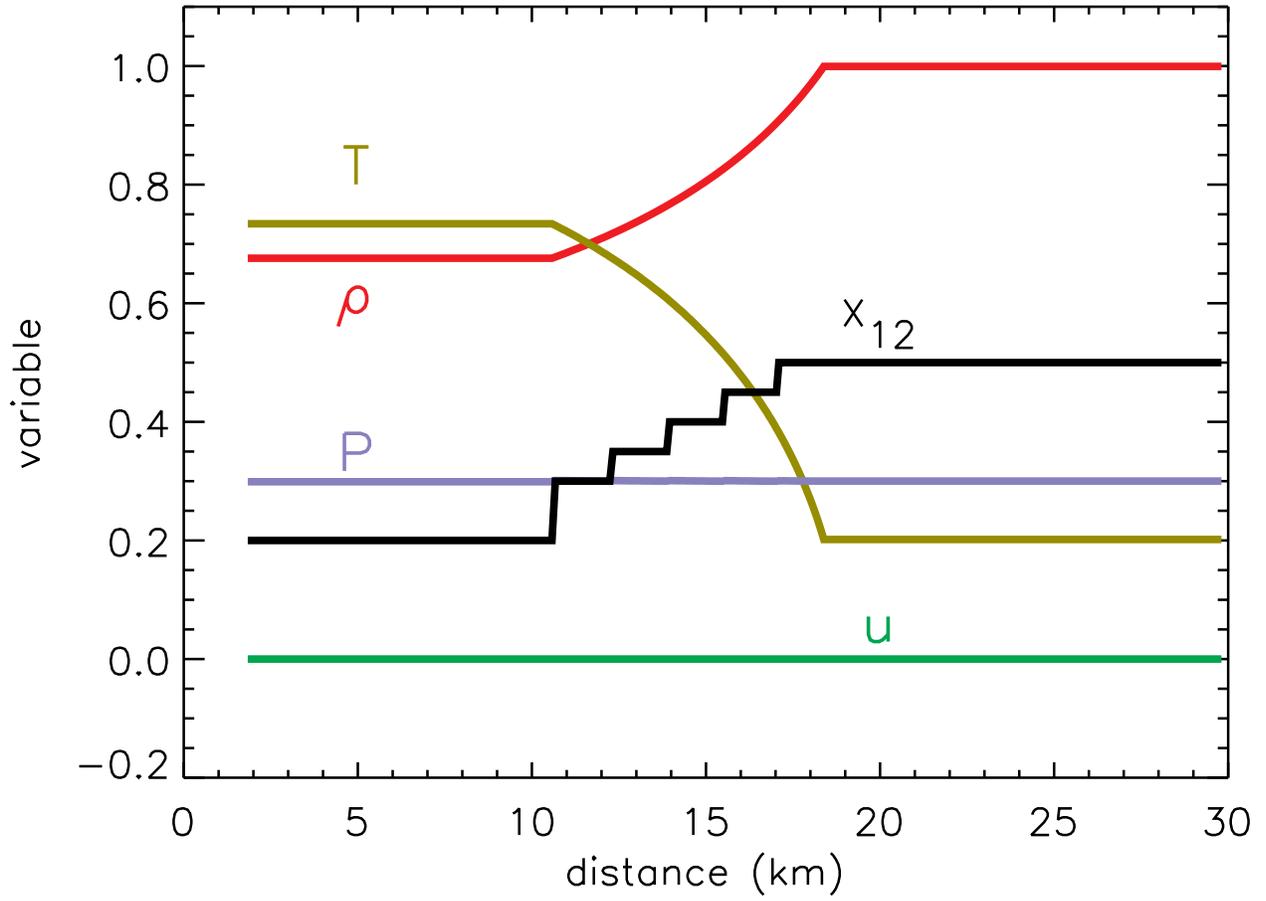}
\caption{Assumed conditions set up by isobaric mixing in a fuel-ash
  mixture with a fuel density of $1 \times 10^7$ g cm$^{-3}$ and a
  carbon mass fraction of 0.50 (see Table 3). The nearly isothermal
  region in the inner 10 km has been heating from burning for about
  0.01 s following mixing that produced a temperature, $T_{\rm
    Max,mix} \approx 1.81 \times 10^9$ K and is now poised to
  runaway.The temperature in the isothermal region is $2.20 \times
  10^9$ K, just slightly less than the $2.26 \times 10^9$ K given
  in Table 3.  The central density is $7 \times 10^6$ g cm$^{-3}$ and
  the carbon mass fraction is 0.20. For purposes of plotting the
  temperature has been divided by $3 \times 10^9$ K, the density, by
  $1 \times 10^7$ g cm$^{-3}$ and the pressure, by $3 \times 10^{24}$
  dyne cm$^{-2}$. \lFig{rsoninit}}
\end{figure}

\begin{figure}
\includegraphics[angle=90,width=\columnwidth]{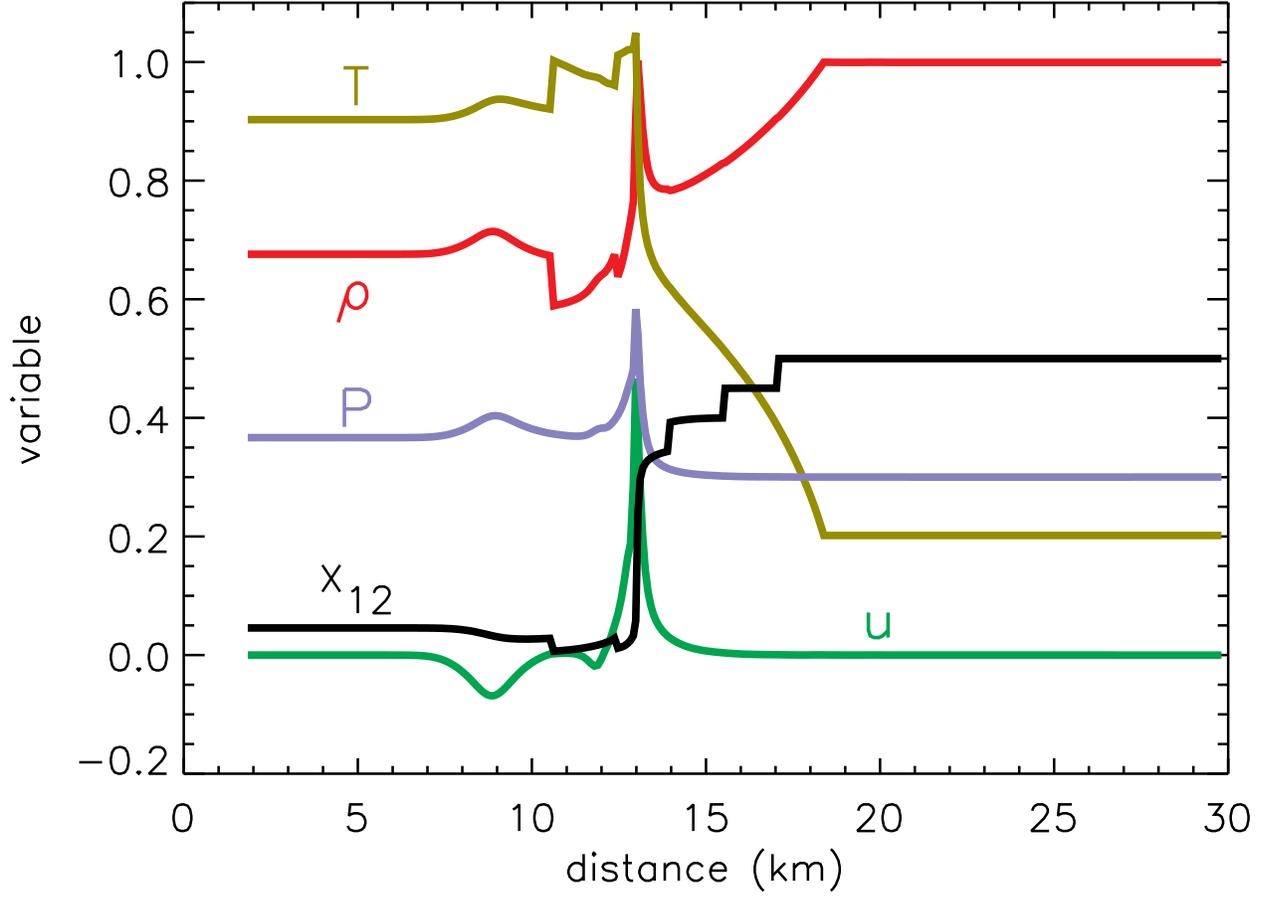}
\caption{The evolution of the conditions shown in \Fig{rsoninit} at a
  time 0.68 ms later. The carbon in the isothermal region has mostly
  burned away and a detonation is forming in the surrounding
  temperature and carbon gradient. This detonation was followed in a
  separate calculation for over 100 km and rapidly grew in strength to
  consume both oxygen and carbon. The variables here are scaled by the
  same amounts as in \Fig{rsoninit}. The velocity, u, is divided by
  4000 km s$^{-1}$ \lFig{rsondet}}
\end{figure}

\begin{deluxetable}{ccccccc} 
\tablecaption{Turbulent flame properties}
\tablehead{
X$_{12}$ & $\rho_{\rm fuel}$ & u$_{L}$ at 10 km & T$_{ash}$ & $\rho_{\rm ash}$ & $\lambda$ & $v_T$ \\
        & (10$^7$ g cm$^{-3}$) & (10$^7$ cm s$^{-1}$) & (10$^9$ K) & (10$^7$ g cm$^{-3}$ & (cm) & (cm s$^{-1}$)  }
\startdata
0.5  & 0.6 &  1 & 2.45 & 0.284 & [1.18(6)] & 1.0(7)     \\
0.5  & 0.6 &  3 & 2.45 & 0.284 & [6.15(6)] & [5.5(7)]   \\
0.5  & 0.8 &  3 & 2.64 & 0.400 & 5.19(5)   & 2.4(7)     \\
0.5  & 0.8 & 10 & 2.64 & 0.400 & [3.16(6)] & [1.5(8)]   \\
0.5  & 1   &  1 & 2.79 & 0.523 & 1.55(4)   & 2.5(6)     \\
0.5  & 1   &  3 & 2.79 & 0.523 & 8.30(4)   & 1.3(7)     \\
0.5  & 1   & 10 & 2.79 & 0.523 & 4.92(5)   & 7.9(7)     \\
0.5  & 2   &  1 & 3.30 & 1.18  & 6.82(1)   &  4.1(5)    \\
0.5  & 2   &  3 & 3.30 & 1.18  & 3.54(2)   &  2.1(6)    \\
0.5  & 2   & 10 & 3.30 & 1.18  & 2.16(3)   &  1.3(7)    \\
0.5  & 2.3 &  1 & 3.40 & 1.39  & 2.43(1)   &  2.9(5)    \\
0.5  & 3   &  1 & 3.63 & 1.89  & 3.62(0)   &  1.5(5)    \\
0.5  & 3   & 10 & 3.63 & 1.89  & 1.15(2)   &  4.9(6)    \\
0.75 & 0.6 &  1 & 2.68 & 0.23  & 3.85(4)   &  3.4(6)    \\
0.75 & 0.6 & 10 & 2.68 & 0.23  & 1.22(6)   &  1.1(8)    \\
0.75 & 0.8 &  1 & 2.89 & 0.33  & 2.99(3)   &  1.4(6)    \\
0.75 & 0.8 & 10 & 2.89 & 0.33  & 9.44(4)   &  4.6(7)    \\
0.75 & 1   &  1 & 3.07 & 0.43  & 4.38(2)   &  7.6(5)    \\
0.75 & 1   & 10 & 3.07 & 0.43  & 1.38(4)   &  2.4(7)    \\
0.75 & 2   &  1 & 3.69 & 1.00  & 1.60(0)   &  1.2(5)    \\
0.75 & 2   & 10 & 3.69 & 1.00  & 5.07(1)   &  3.7(6)    \\
0.75 & 3   & 10 & 4.09 & 1.62  & 2.47(0)   &  1.3(6)    \\
\enddata
\end{deluxetable}

\begin{deluxetable}{cccccc} 
\tablecaption{Ash temperature}
\tablehead{
$\rho_7$ & X$_{12}$ = 0.50 & X$_{12}$ = 0.75 & $\rho_7$ & X$_{12}$ = 0.50 & X$_{12}$ = 0.75 }
\startdata
0.6 & 2.45 & 2.68 & 2.0 & 3.30 & 3.69  \\
0.8 & 2.64 & 2.89 & 2.2 & 3.37 & 3.78  \\
1.0 & 2.79 & 3.07 & 2.4 & 3.44 & 3.86  \\
1.2 & 2.92 & 3.22 & 2.6 & 3.51 & 3.94  \\
1.4 & 3.03 & 3.36 & 2.8 & 3.57 & 4.02  \\
1.6 & 3.13 & 3.48 & 3.0 & 3.63 & 4.09  \\
1.8 & 3.22 & 3.59 & 3.2 & 3.68 & 4.16  \\    
\enddata
\end{deluxetable}

\begin{deluxetable}{cccccccccc} 
\tablecaption{Conditions for supersonic burning}
\tablehead{
X$_{12}$ & $\rho_7$ & X$_{12,0.02}$ & T$_{0.02}$ & $\rho_{0.02}$ &  X$_{12,max}$ & 
T$_{\rm max}$ &  $\rho_{\rm max}$ & $\tau_{\rm nuc,min}$ & $r_{\rm sonic}^{\rm min}$ \\
        & (10$^7$ g cm$^{-3}$) &  & (10$^9$ K) & (10$^7$ g cm$^{-3}$ & 
& (10$^9$ K) & (10$^7$ g cm$^{-3}$ & (sec) & (cm)}
\startdata
0.5  & 0.8  & 0.26 & 2.09  & 0.53 & 0.20  & 2.22 & 0.50 & 1.7(-3) & 7.5(5)   \\
0.5  & 1.0  & 0.36 & 1.77  & 0.78 & 0.20  & 2.26 & 0.66 & 7.7(-4) & 3.5(5)   \\
0.5  & 1.5  & 0.40 & 1.65  & 1.27 & 0.20  & 2.43 & 0.45 & 1.1(-4) & 5.4(4)   \\
0.5  & 2.0  & 0.41 & 1.60  & 1.76 & 0.20  & 2.56 & 1.46 & 2.9(-5) & 1.4(4)   \\
0.5  & 2.5  & 0.43 & 1.56  & 2.26 & 0.20  & 2.67 & 1.87 & 1.0(-5) & 5.4(3)   \\
0.5  & 3.0  & 0.43 & 1.54  & 2.75 & 0.20  & 2.77 & 2.12 & 4.8(-6) & 2.5(3)   \\
     &      &      &       &      &       &      &      &         &          \\
0.5  & 0.8  & 0.26 & 2.09  & 0.53 & 0.13  & 2.36 & 0.47 & 1.2(-3) & 5.7(5)   \\
0.5  & 1.0  & 0.36 & 1.77  & 0.78 & 0.12  & 2.46 & 0.61 & 4.7(-4) & 2.2(5)   \\
0.5  & 1.5  & 0.40 & 1.65  & 1.27 & 0.11  & 2.70 & 0.97 & 6.0(-5) & 3.0(4)   \\
0.5  & 2.0  & 0.41 & 1.60  & 1.76 & 0.10  & 2.88 & 1.35 & 1.5(-5) & 7.4(3)   \\
0.5  & 2.5  & 0.43 & 1.56  & 2.26 & 0.10  & 3.02 & 1.73 & 5.1(-6) & 2.7(3)   \\
0.5  & 3.0  & 0.43 & 1.54  & 2.75 & 0.098 & 3.14 & 2.04 & 2.4(-6) & 1.3(3)  \\
     &      &      &       &      &       &      &      &         &          \\
0.75 & 0.6  & 0.43 & 2.17  & 0.34 & 0.30  & 2.36 & 0.30 & 6.9(-4) & 3.3(5)   \\
0.75 & 0.8  & 0.61 & 1.68  & 0.61 & 0.30  & 2.47 & 0.44 & 1.8(-4) & 8.6(4)   \\
0.75 & 1.0  & 0.63 & 1.62  & 0.81 & 0.30  & 2.60 & 0.58 & 5.2(-5) & 2.6(4)   \\
0.75 & 1.5  & 0.66 & 1.55  & 1.30 & 0.30  & 2.83 & 0.93 & 6.4(-6) & 3.3(3)   \\
0.75 & 2.0  & 0.67 & 1.52  & 1.78 & 0.30  & 3.01 & 1.29 & 1.5(-6) & 8.2(2)   \\
0.75 & 2.5  & 0.68 & 1.49  & 2.27 & 0.30  & 3.15 & 1.67 & 5.e(-7) & 3.0(2)   \\
0.75 & 3.0  & 0.69 & 1.47  & 2.77 & 0.30  & 3.26 & 2.06 & 2.4(-7) & 1.3(2)   \\
     &      &      &       &      &       &      &      &         &          \\
0.75 & 0.6  & 0.43 & 2.17  & 0.34 & 0.24  & 2.43 & 0.29 & 6.3(-4) & 3.1(5)   \\
0.75 & 0.8  & 0.61 & 1.68  & 0.61 & 0.21  & 2.60 & 0.41 & 1.4(-4) & 7.1(4)   \\ 
0.75 & 1.0  & 0.63 & 1.62  & 0.81 & 0.21  & 2.75 & 0.53 & 4.2(-5) & 2.1(4)   \\
0.75 & 1.5  & 0.66 & 1.56  & 1.30 & 0.19  & 3.04 & 0.85 & 4.7(-6) & 2.5(3)   \\
0.75 & 2.0  & 0.67 & 1.51  & 1.79 & 0.19  & 3.27 & 1.19 & 1.1(-6) & 5.8(2)   \\
0.75 & 2.5  & 0.68 & 1.49  & 2.27 & 0.18  & 3.45 & 1.53 & 3.6(-7) & 2.0(2)   \\
0.75 & 3.0  & 0.68 & 1.46  & 2.77 & 0.17  & 3.60 & 1.89 & 1.5(-7) & 8.5(1)   \\
\enddata
\end{deluxetable}

\begin{deluxetable}{cccccccc} 
\tablecaption{Detonations}
\tablehead{
X$_{12}$ & $\rho_7$ & X$_{12,1}$ & $T_1$ & $\rho_1$ & $L_1$ & $L_2$ & detonates?}
\startdata
0.5 & 0.8 & 0.20 & 2.22 & 0.50 & 1.4(6) & 1.4(6)  & no \\
0.5 & 0.8 & 0.20 & 2.22 & 0.50 & 2.8(6) & 2.8(6)  & yes \\
0.5 & 1.0 & 0.20 & 2.26 & 0.66 & 5.2(5)  & 4.4(5)  &  no   \\
0.5 & 1.0 & 0.20 & 2.26 & 0.66 & 1.1(6)  & 9.4(5)  &  yes   \\
0.75 & 0.6 & 0.30 & 2.36 & 0.30 & 5.5(5) & 5.0(5)  &  no  \\
0.75 & 0.6 & 0.30 & 2.36 & 0.30 & 1.1(6) & 1.0(6)  &  yes \\
0.75 & 0.8 & 0.30 & 2.47 & 0.44 & 1.0(5) &  7.8(4) & no   \\
0.75 & 0.8 & 0.30 & 2.47 & 0.44 & 2.2(5) &  1.6(5) & yes  \\
0.75 & 1.0 & 0.30 & 2.60 & 0.58 & 2.8(4) &  2.2(4) & no   \\
0.75 & 1.0 & 0.30 & 2.60 & 0.58 & 6.1(4) &  4.8(4) & yes  \\
\enddata
\end{deluxetable}

\end{document}